\documentclass{osa-article}
\usepackage{filecontents}
\journal{osac}

\usepackage[utf8]{inputenc}
\usepackage{multicol}
\usepackage{xcolor}
\usepackage{braket}
\usepackage{nicefrac}
\usepackage{wrapfig}
\usepackage{bm}
\def\ve#1{\bm{\mathbf{#1}}}
\articletype{Research Article}

\usepackage{lineno}

\begin{document}

\title{Coherent Stokes Raman scattering microscopy (CSRS)}

\author{Sandro Heuke\authormark{1,*} and Hervé Rigneault\authormark{1,*}}

\address{\authormark{1}Aix Marseille Univ, CNRS, Centrale Marseille, Turing Center for Living Systems, Institut Fresnel, Marseille, France.}
\email{\authormark{*}Corresponding authors: Sandro.Heuke@fresnel.fr \& herve.rigneault@fresnel.fr} 



\begin{abstract}
We report the first implementation of laser scanning Coherent Stokes Raman scattering (CSRS - pronounced "sCiSsoRS") microscopy. 
To overcome the major challenge in CSRS imaging, we show how to suppress the fluorescence background by narrow bandpass filter and a lock-in based demodulation. Near background free CSRS imaging of polymer beads, human skin, onion cells, avocado flesh and the wing disc of a drosphila larva are presented. Finally, we explain and demonstrate numerically that CSRS solves a major obstacle of other coherent Raman techniques by sending a significant part (up to 100\%) of the CSRS photons into the backward direction under tight focusing conditions. We believe that this discovery will pave the way for numerous technological advances, e.g. in epi-detected coherent Raman multi-focus imaging, real-time laser scanning based spectroscopy or efficient endoscopy.                 
\end{abstract}

\section{Introduction}
\begin{figure}[htbp]
\centering
\fbox{\includegraphics[width=11cm]{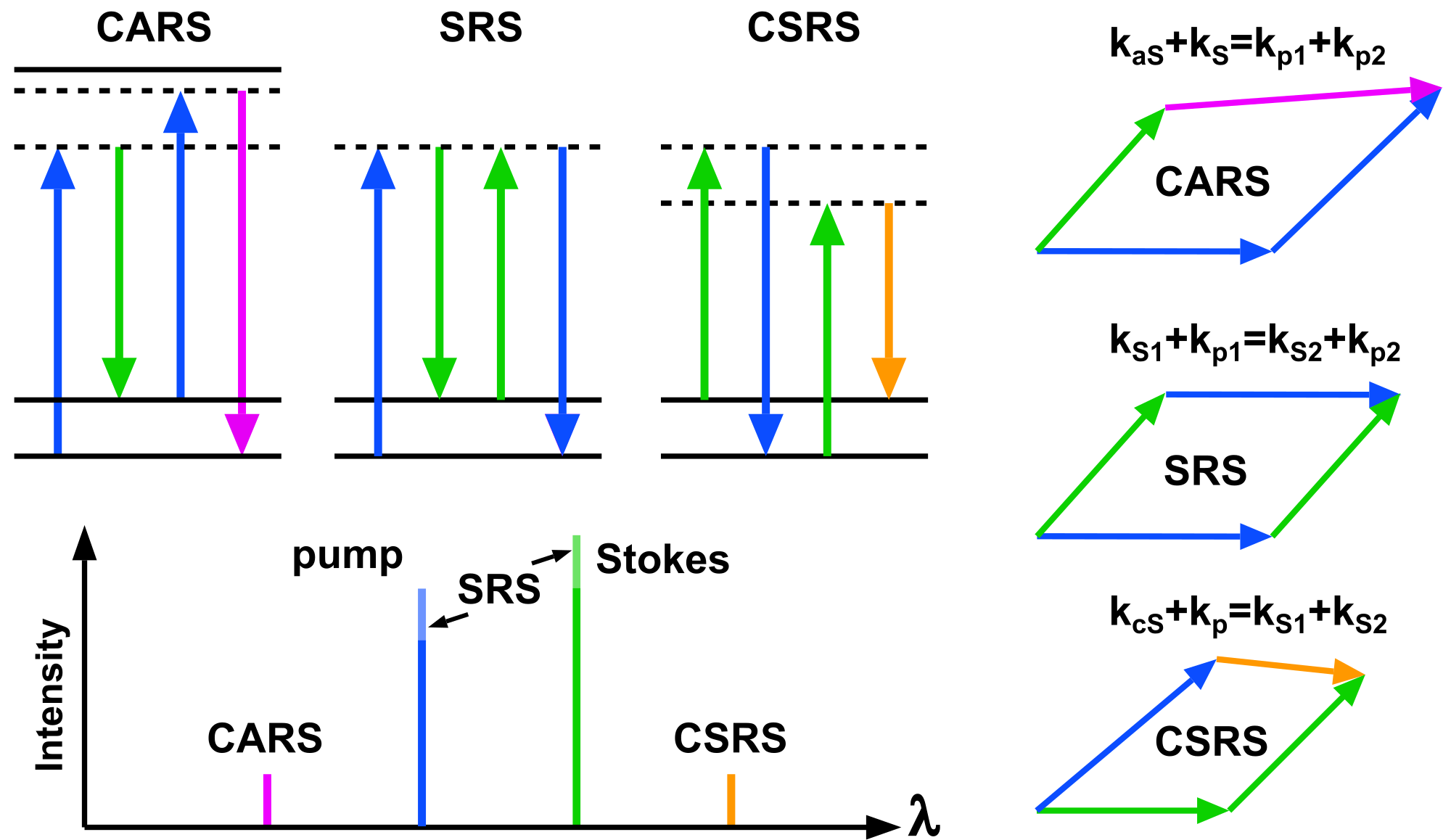}}
\caption{Coherent Raman imaging techniques in energy diagrams, relative radiation wavelength and energy conservation under plane-wave illumination.}
\label{fig:intro}
\end{figure}

Conventional bright-field microscopy provides information about the refractive index and absorption properties, but cannot elucidate the sample's chemical composition. Infra-red absorption and linear Raman scattering retrieve the chemical fingerprint \cite{Bunaciu2013,Antonio2013}, but are incompatible with high spatial resolution or real-time imaging. 
Coherent Raman imaging (CRI) fills this technological gab joining a chemical bond specific contrast with signal levels that permit video-rate image acquisition. 
Well established CRI microscopy techniques are the coherent anti-Stokes Raman scattering (CARS) \cite{Duncan1982, Zumbusch1999} and stimulated Raman scattering (SRS) \cite{Nandakumar2009, Freudiger2008, Ozeki2009}. CARS owes its wide-range application to the blue-shifted anti-Stokes radiation which greatly facilitates its separation from linear fluorescence. 
When working with near infra-red excitation wavelength, the blue-shifted CARS radiation is readily detected using photo-election multiplier tubes (PMT) of standard laser scanning microscopes.
SRS's popularity arises from the homodyne signal amplification that frees SRS images from an omnipresent non-resonant four-wave-mixing background and allows for measurements under daylight conditions.\\
Overshadowed by CARS and SRS until now, there exists a 3$^{\textrm{rd}}$ four-wave-mixing process termed coherent Stokes Raman scattering (CSRS, "Scissors") \cite{Maker1965,Zheng1983,Cui2009} which is always appearing within any CARS or SRS experiment and provides near identical mapping of molecular oscillators \cite{Druet1981} - see Fig.\ref{fig:intro}. In analogy to the Stokes emission in linear Raman microscopy, the CSRS radiation ($2\omega_S-\omega_p$) is red-shifted with respect to the excitation frequencies of the pump ($\omega_p$) and Stokes beams ($\omega_S$). 
Surprisingly, CSRS was not yet implemented for laser scanning microscopy. Presumably, this neglect must be attributed to the high degree of resemblance of CARS and CSRS spectra \cite{Druet1981} rendering CSRS - prima facie - to be either CARS with an added fluorescence background when working with visible light sources or, using near infra-red (NIR) excitation, CARS with a radiation wavelength offside high quantum yields of common detectors.
CSRS provides, however, some unique properties that are of high interest for imaging. 
(1) The CSRS spectrum differs from CARS in the presence of accessible electronic resonances. For example, pre-resonant CSRS will offer complementary information in application to alkyne-labeled dyes \cite{Wei2017} and standard dyes used in microbiology \cite{Wei2018}. (2) The red-shifted radiation of CSRS imaging becomes an advantage for UV or near-UV excitation where CARS photons \cite{Prince2019} would be too far blue-shifted to be detected efficiently while any SRS image \cite{Bi2018} is likely to be compromised by various artifacts such as multi-photon absorption\cite{Berto2014,Heuke2020a}. Thus, UV excited CSRS holds the potential to achieve the highest possible spatial resolution ($\lambda_{\textrm{Stokes}}/[\sqrt{8}NA]$) in coherent Raman imaging. (3) NIR-excitation wavelength combined with CSRS may allow for deeper tissue imaging due to the reduced scattering and absorption of its radiation \cite{Kobat2009}. (4) Last but most important: Due to a modified phase-matching geometry, CSRS microscopy can be configured to radiate more light into the backward direction which will add game-changing benefits for the investigation of thick samples, real-time spectroscopy, multi-focus imaging and endoscopy \cite{Lombardini2018}.
Within this contribution, we want to open up the field of laser scanning CSRS imaging by demonstrating CSRS microscopy within the visible excitation spectrum. To remove the major obstacle, we will show how linear fluorescence can be suppressed by a set of bandpass filter and nearly nullified in combination with a lock-in based detection scheme as a premise for near-UV excited CSRS imaging with a lateral resolution < 100 nm.
Furthermore, we shall investigate numerically CSRS' spatial radiation behavior under NIR excitation paving the way towards CSRS experiments with an efficient epi-detection.

\section{Experimental result and discussion}
\begin{figure}[htbp]
\centering
\fbox{\includegraphics[width=\linewidth]{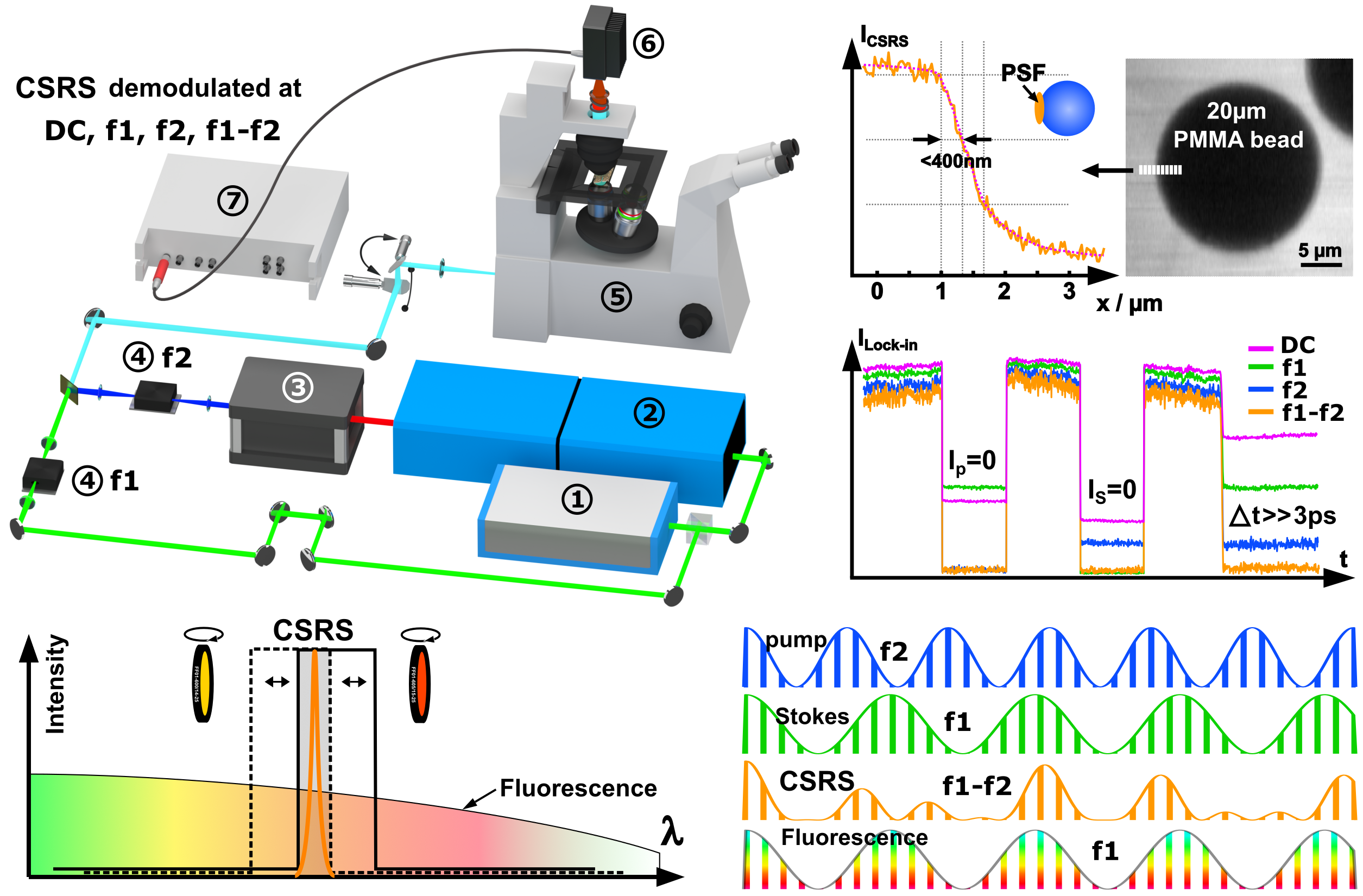}}
\caption{CSRS experimental implementation and characterization. Bottom left: The CSRS signal is separated from fluorescence by means of 2 angle-tuned narrow bandpass filter. Bottom right: Additional suppression of fluorescence is achieved by intensity modulating the Stokes and pump beam at the radio frequencies f1 and f2, respectively. Fluorescence free CSRS signal is obtained at f1-f2. Right center: Time separation of the pump and Stokes pulses as well as blocking the excitation highlights the superior suppression of fluoresence background at the demodulation frequency f1-f2 compared to CSRS signal obtained at f1, f2 or the DC frequency. Top right: The intensity profile at the interface of a PMMA bead and olive oil indicates a lateral resolution of <400nm. Top left: scheme of the CSRS experiment. 1 Yb-fiber laser, 2 optical parametric oscillator (OPO), 3 Second harmonic generation (SHG), 4 acousto-optic modulator (AOM), Laser scanning microscope (LSM), 6 photo-electron multiplier (PMT), 7 Lock-in amplifier.}
\label{fig:experiment}
\end{figure}
\begin{figure}[htbp]
\centering
\fbox{\includegraphics[width=\linewidth]{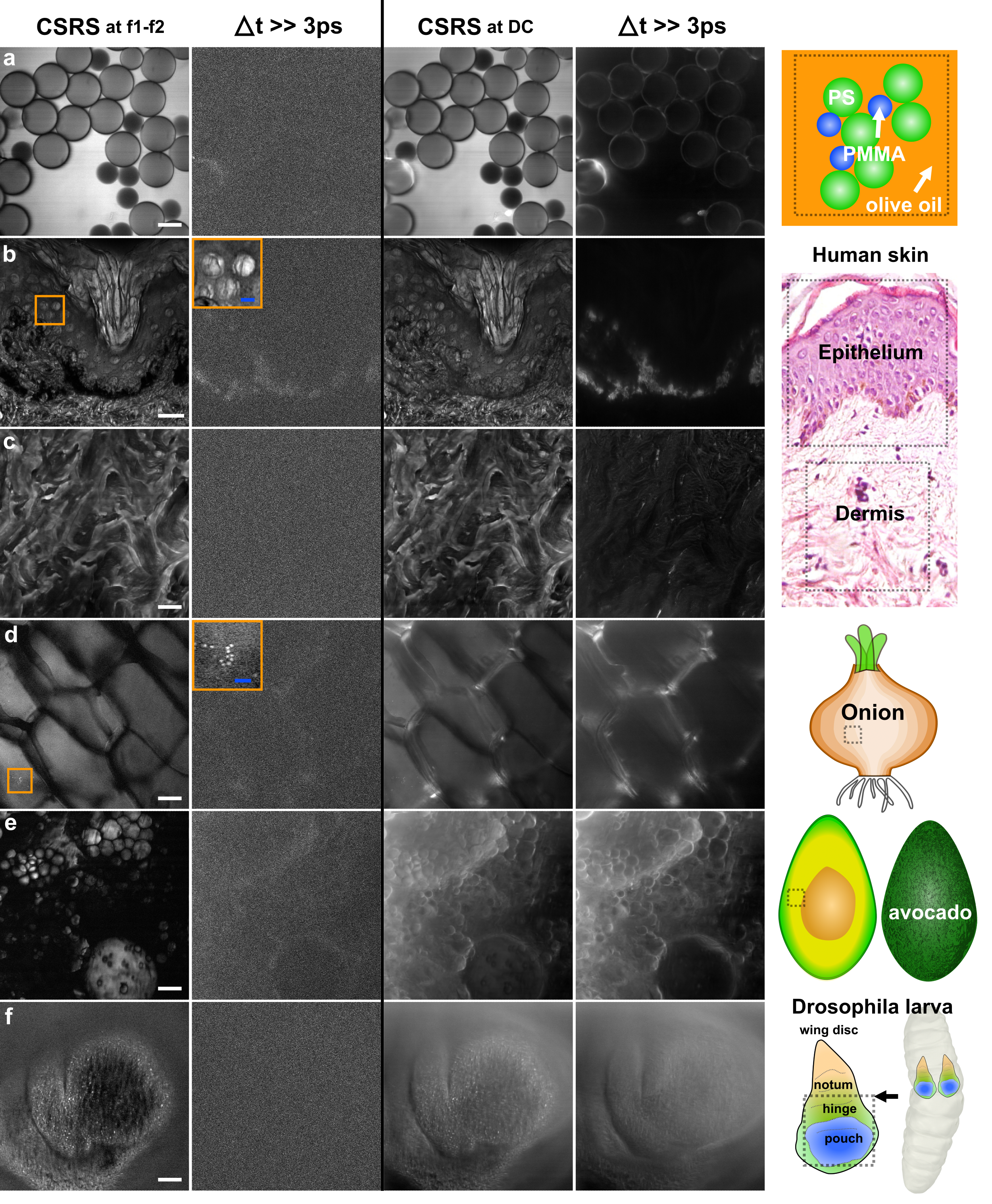}}
\caption{LSM-CSRS at 2850 cm$^{-1}$. The left and right column show the CSRS image demodulated at the frequencies f1-f2 = 1.47 MHz and 0 Hz (DC). To estimate the remaining fluorescence level, images without temporal overlap of the pump and Stokes pulses are displayed to the right. a) Mixture of polystyrene (PS, 30µm) and Poly-methyl-methacrylate (PMMA, 20µm) beads in olive oil. b) and c) Epithelium and dermis of a 20µm thick human skin section d) Cells of an onion. e) Lipid droplets within the flesh of an avocado. d) Wing disc of a Drosophila larva. The white and blue scale bar equals 20µm and 5µm, respectively.}
\label{fig:images}
\end{figure}
The CSRS signal of biomedical samples is readily overwhelmed by linear fluorescence. Time-gating \cite{Koegler2020}, a time-resolved detection using streak cameras \cite{Tahara1993} or polarization filtration can be used to reduce or suppress any fluorescence signal. These methods require, however, either a substantial alteration of standard coherent Raman microscopes or do not work in the presence of large quantities of fluorescence light. Here, we exploit the fact that the CSRS is spectrally narrow under ps-excitation. Thus, the majority of fluorescence is readily suppressed by the choice of a the narrow-band filter. Filters with a spectral width below < 1nm are commercially available but the selection of a specific center wavelength requires expensive costume solutions. This is the reason why we use a combination of two inexpensive bandpass filter with a width of about 15 nm, but different center wavelength. In a addition, we fine-tune the filter transmission by a tilt (<20$^{\circ}$) with respect to the incident beam. Thus, two tilt-adjusted bandfilter create a sharp transmission line (FWHM<3nm) for the CSRS signal while rejecting significant parts of the autofluorescence.\\
As a second method for fluorescence discrimination, we take advantage of CSRS intensity dependence on both excitation colors while linear fluorescence follows either the intensity of the pump or the Stokes laser. Consequently, modulating the pump and Stokes beams at f1 and f2 while demodulation the signal at f1-f2 (or f1+f2) yields exclusively nonlinear signals that depend on both excitation colors. The f1-f2 demodulation, therefore, also discriminates the CSRS signal against 2-photon excited fluorescence (2PEF) under single-color excitation. It shall be noted that the double modulation is also sensitive to two-color 2-photon fluorescence (2C-2PEF). Nevertheless, we will find experimentally, that the emission strength of native 2C-2PEF is negligible within our CSRS approach. \\
For the experimental implementation of CSRS into laser scanning microscopy, we chose visible excitation wavelengths at 445nm (pump) and 515nm (Stokes) for the following reasons: (1) CSRS under near UV excitation is a potentially important application area since the CARS signal falls into the UV range while SRS artifacts are increased due the high concentration of matching chromophores. (2) The red-shifted CSRS radiation is readily detected by ordinary PMTs. (3) Fluorescence artifacts are enhance compared to a near infra-red (NIR) excitation. Thus, our approach will be viable as well for CSRS under NIR excitation, if pure CSRS signals can be obtained under VIS excitation. The experimental implementation, the spectral filtration and the double modulation are schematically shown in Fig.\ref{fig:experiment}a. Our implementation resembles a standard SRS setup with the difference that we use visible excitation wavelengths, we modulate not one but both beams and the photo-diode is replaced by a PMT which is connected to a lock-in amplifier. More information about the setup can be found within the part Methods: Experimental setup. To quantify the level of fluorescence rejection, we investigated the signal of native olive oil at 2850~cm$^{-1}$ when blocking the Stokes or pump beams and when the temporal pulse overlap is removed. The output signal of the lock-in is plotted as functions of the demodulation frequencies at 0 Hz (DC), f1, f2 and f1-f2 in Fig.\ref{fig:experiment}. It can be observed that the DC channel contains significant amounts of fluorescence while this artifact is already reduced within the channels f1 and f2. Nevertheless, only the difference frequency channel at f1-f2 becomes dark, when the excitation pulses do not overlap in time.  In a second experiment, we imaged the interface of olive oil and a 20µm sized Plexiglas (PMMA) bead  to obtain an estimation of the lateral resolution for an excitation objective featuring an NA of 1.45 - see Fig.\ref{fig:experiment}. From this "knife-edge" CSRS intensity profile, we can infer a lateral resolution below 400nm. The difference to the expected $\lambda_{Stokes}/[\sqrt{8}NA]$= 515nm/[$\sqrt{8}$1.49]=120nm can be attributed to underfilling of the excitation objective lens and the bent oil/bead interface. Having confirmed a high-resolved, fluorescence-free CSRS image contrast, we investigated the suitability of LSM-CSRS for vibrational imaging of various objects featuring non-negligible fluorescence levels. Within Fig.~\ref{fig:images}, we show the CSRS images of test and biomedical samples demodulated at the DC and f1-f2 frequencies for (non-)overlapping pump and Stokes pulses. The images were organized along the ratio of the CSRS to fluorescence signal starting from the highest at the top. Comparing the DC and f1-f2 images in Fig.~\ref{fig:images}a, it obvious that a narrow spectral filtering is already sufficient for CSRS imaging of polymer beads in oil. The first artifacts become visible for the DC CSRS images of the epithelium and dermis of a 20µm thick section of human skin - see Figs.~\ref{fig:images}b and c. For the epithelium, a pronounced fluorescence artifact arises from melanin within the Stratum basale. Artifacts within the Dermis can be attributed to the auto-fluorescence of collagen and elastin \cite{Heuke2013}. The quantity of fluorescence observed within the DC channel increases stepwise further for CSRS imaging of onion cells, lipid droplets within the flesh of an avocado and the wing disc of a Drosophilia larva. From the second row of Fig.~\ref{fig:images}, it is reconfirmed that almost no fluorescence is leaking into the f1-f2 CSRS channel as an important condition for the estimation of the true concentration of the targeted molecular group. The origin of fluorescence for these 3 samples, however, cannot be attributed with certainty, but might arise from NADH, flavins and chlorophyll.  \\
In a broader context, we would like to point out that other nonlinear microscopy techniques would also greatly benefit from the narrow-band filter plus demodulation combination for rejection of spurious background signals. For example, the 2PEF signal of chlorophyll in plant leaves readily overwhelms any CARS or second harmonic generation (SHG) image contrast even under NIR excitation. A double modulation of the excitation combined with a lock-in based demodulation will purify the signal, reduce the sensitivity against other light sources such as room light and reestablish the reliability of the following image analysis. 
Having removed the why-not argument for the CSRS image contrast, we shall introduce in the next section a non-intuitive but game-changing argument for CSRS microscopy : the increased backwards radiation as the prerequisite of an effective epi-CSRS detection.  

\section{Numerical results}
\begin{figure}[htbp]
\centering
\fbox{\includegraphics[width=\linewidth]{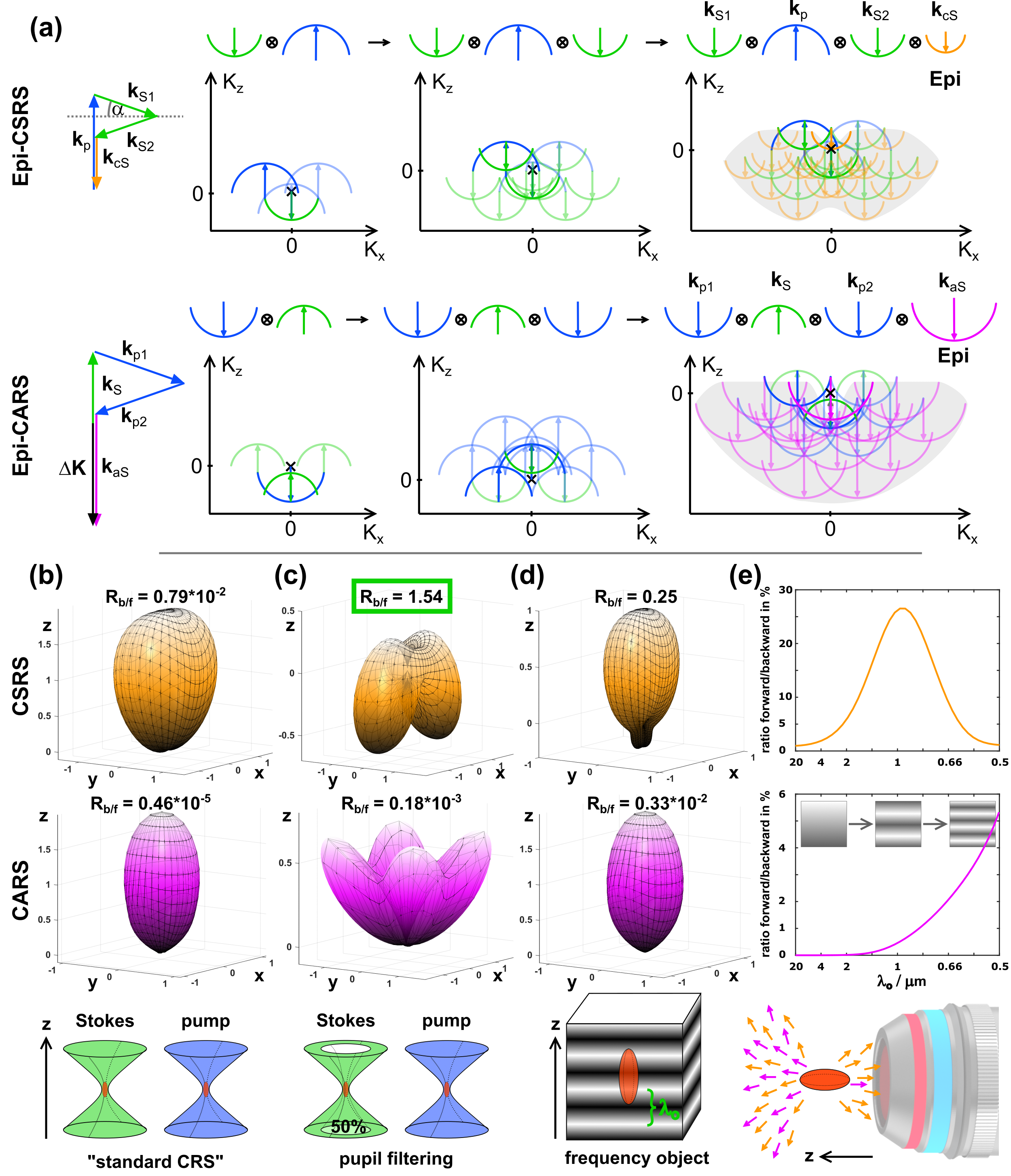}}
\caption{Object frequency support and radiation behavior of CSRS versus CARS. a) The object K-support for Epi-CSRS(CARS) is found by convolving the illumination Ewald spheres of the Stokes (pump), pump (Stokes), and Stokes (probe) with the cap of detection Ewald sphere at (anti-)Stokes frequency. Note that vector combinations covering the frequency of a homogeneous sample K(0,0,0) are only found for CSRS but not for CARS. A single wavevector combination that phase-matches K(0,0,0) is highlighted to the left while a similar approach for CARS leads to a large phase-mismatch ($\Delta$K). b) CSRS and CARS radiation behavior of a homogeneous sample under standard illumination condition, i.e. the pump and Stokes beam fill the objective aperture homogeneously ($\theta_{max}$=80$^\circ$). c) same as in b) but with an annular pupil filter applied to the Stokes beam for CSRS covering 50\% of area of the objective back-aperture. For an equitable comparison with CARS, the same pupil filter was applied to the pump beam. d) same as for b) but the homogeneous sample was replaced by a frequency object whose scatter density is described as $1+\cos(2\pi z/\lambda_o)$ and $\lambda_o$=1µm. e) Plot of the ratio of backward/forward radiation (R$_{\textrm{b/f}}$) as a function of the object frequency $\lambda_o$.}
\label{fig:object_support}
\end{figure}

In this section, we shall show and explain CSRS' superior backward radiation properties. Before entering into the calculations, we want to consider CSRS from a heuristic viewpoint investigating the momentum conservation laws for CSRS and compare it to CARS. Under plane illumination, the momentum conservation laws can be written as $\ve{K}=\ve{k}_{p}-\ve{k}_{S}+\ve{k}_{p}-\ve{k}_{aS}$ for CARS \cite{Heuke2019} and $\ve{K}=\ve{k}_{S}-\ve{k}_{p}+\ve{k}_{S}-\ve{k}_{cS}$ for CSRS with $\ve{K}$, $\ve{k}_{p}$, $\ve{k}_{S}$, $\ve{k}_{aS}$ and $\ve{k}_{cS}$ representing the wavevectors of the object, the pump(probe) and Stokes beam as well as the anti-Stokes and coherent Stokes radiation, respectively. Note that for homogeneous samples these laws are also referred to as phase-matching condition and simplify to $\ve{k}_{p}+\ve{k}_{p}=\ve{k}_{S}+\ve{k}_{aS}$ (CARS) and $\ve{k}_{S}+\ve{k}_{S}=\ve{k}_{p}+\ve{k}_{cS}$ (CSRS). Under focusing conditions, the single wavevectors are replaced by the distribution of incident wavevectors which are distributed over a cap of a sphere. To identify those object frequencies (\ve{K}) that are effectively probed, every combination of excitation and emission wavevector must be identified. This operation is equivalent to the convolution of the caps of the illumination and detection Ewald spheres. Neglecting polarization effects, the result of this convolution (simplified to 3 points per arc) is shown in 2D within Fig.~\ref{fig:object_support}a. 

Evidently, there exist no vector combination for epi-scattered CARS photons which would cover the origin K(0,0,0) of the object space. Thus, a homogeneous sample, such as olive oil, does not provide any backward radiation. On the contrary, structures that feature high object frequencies, such as small polymer beads or layered materials, generate Epi-CARS radiation. In the past, Epi-CARS was occasionally considered to be a size selective contrast that would highlight exclusively small objects \cite{Volkmer2001}. While this statement holds for the majority of biomedical samples, there do exist large structures, e.g. multi-layered lipids in vesicles that also emit a strong CARS radiation into the backward direction. Hence, it is more appropriate to refer to Epi-CARS as a technique that probes high object frequencies instead of been considered as size selective.\\
Switching the detection wavelength to the red-shifted coherent Stokes radiation changes the covered object support significantly and includes now the origin at K(0,0,0). Due to the reduced size of the detection wavevector (|\ve{k}$_{cS}|\ll|$\ve{k}$_{aS}$|) and the pump vector entering as complex conjugated, see Eq.~\ref{eq.:P_aS}, it is now possible to find vector combinations that cover the origin at K(0,0,0). Consequently, even a homogeneous object will radiate considerable amounts of Epi-CSRS. Nevertheless, since the the centroid of the Epi-CSRS object support, i.e. the gray cloud within Fig.~\ref{fig:object_support}a, does not coincidence with the K-space origin, Epi-CSRS images will also highlight objects containing higher frequencies.\\
To address the question of how to increase the ratio of Epi versus forward Epi-CSRS, and which object frequencies are most efficiently probed using Epi-CSRS, we performed finite element simulations whose results are summarized in Fig.~\ref{fig:object_support}b-e. The equations implemented numerically as well as important parameters are found in the annex - numerical calculation. From the momentum conservation law and the vector diagrams in Fig.~\ref{fig:object_support}a, it is readily comprehensible that a larger wavelength difference in between the pump and coherent Stokes wavelength relaxes greatly the necessity for extreme incident illumination angles of the Stokes beam. Furthermore, since most of the coherent Raman experiments apply NIR instead of VIS excitation wavelength, we used for within our simulations the wavelength $\lambda_p=797nm$ and $\lambda_S=1030nm$ which matches the most commonly targeted Raman shift in CRI imaging at 2850cm$^{-1}$. For these conditions, the coherent Stokes radiation will be observed at $\lambda_{cS}=1450nm$. It shall be noted that our results equally apply for the visible excitation wavelength with gently higher excitation angle or thinner annular masks. \\  
To start with, we computed the radiation pattern of CSRS and CARS of a homogeneous object using an NA of 1.49 (oil immersion) corresponding to a maximum illumination angle of 80$^\circ$. From Fig.~\ref{fig:object_support}b, it is evident that both CARS and CSRS are predominately forward directed though the CSRS' radiation distribution features a larger radiation cone. Considering the ratio of backward versus forward directed photons R$_{\textrm b/f}$, we find numerically that less than 1 photon in 10$^5$ is backward directed for CARS. Note that the momentum conservation actually law predicts R$_{\textrm b/f}$=0 for CARS. Thus, the resulting deviation must be attributed to the finite number of voxels of the numerical model. 
For CSRS, R$_{\textrm b/f}$ increase dramatically to about 1 in 100 photons.
Since common surfaces within biomedical samples scatter more than 1\%, we have to assume, however, that also epi detected CSRS will be just forward generated CSRS that was redirected by linear scattering at an interface. Still, using a confocal detection, i.e. a pinhole in front of the detector placed at the conjugated plane of the excitation focus, might already yield true Epi-CSRS images of homogeneous samples where Epi-CARS images would remain dark. To find an approach that increases the proportion of CSRS' epi radiation, we shall consider the CSRS vector diagram matching K(0,0,0) on the left of Fig.~\ref{fig:object_support}a. The ratio of backward versus forward radiation is readily increased by reducing the impact of vectors combinations probing higher frequencies and favoring those covering the origin. This boost of epi-CSRS radiation can be achieved using an annular illumination of the Stokes beam. Experimentally, such an annular illumination is generated, without power-loss, using 2 axicons within the Stokes beam path \cite{Heuke2015a,Heuke2015b}.   
Numerically, we restricted the incident angles for the Stokes between $\theta_{min}$=56.5$^{\circ}$ and $\theta_{max}$=80$^{\circ}$, which corresponds to covering 50\% of the area of the objective lens' back-focal plane. With this pupil filtering, the radio of backward to forward radiation increased for CARS to 2 in 10$^4$ photons while the majority of all CSRS radiation is backward directed (R$_{\textrm b/f}$=1.5) when focusing the pump and Stokes beam into a homogeneous object - see Fig.~\ref{fig:object_support}c.\\
As a second important result from the heuristic derivation of CSRS' object support, we found that the presence of high object frequencies increases the amount of backward radiation. To confirm this prediction, we investigated in Fig.~\ref{fig:object_support}d and e an object whose nonlinear scatterer density, i.e. concentration of molecular groups, is modulated along the optical axis as 1+$\cos(K_zz)$ with $K_z=2\pi/\lambda_o$ being the object frequency.  
As an example, Fig.~\ref{fig:object_support}d outlines the radiation behavior of a wave-like structured object with K$_z$=2$\pi$/1µm. It is found that R$_{b/f}$ increases to one forth for Epi-CSRS while Epi-CARS remains negligible weak. 
To identify those object frequencies which are most efficiently probed by Epi-CSRS, we computed R$_{b/f}$ as a function of K$_z$. From Fig.~\ref{fig:object_support}e, we find that Epi-CSRS peaks at K$_z$=2$\pi$/1µm whereas Epi-CARS R$_{b/f}$ still increases at $K_z=2\pi/0.25$µm. 
It shall be note that the R$_{\textrm b/f}$ never reaches 1 which arise from the 1+ within the definition of the frequency object (1+$\cos(K_zz)$). The 1+ implies that the wave-object always features twice the amplitude at K(0,0,0), which corresponds to a homogeneous predominantly forward scattering object, compared to the scatterer density modulation K(0,0,$\pm$K$_z$). Our simulation results in a nutshell: we have found that CSRS features a non-negligible backward radiation from a homogenous sample under tight-focusing conditions while this is not the case for CARS. The amount of backward radiated CSRS can be enhanced by increasing the illumination power of the Stokes beam with high incident angles. Furthermore, the natural structure of biomedical samples, which are usually not homogeneous, will also elevate the CSRS backward radiation.    
\section*{Conclusion}
We have demonstrated the first laser scanning microscopy CSRS experiment. As the major challenge, we were able to reduce the fluorescence background significantly using a pair of tilted bandpass filter. The remaining fluorescence contribution is removed by intensity modulating the Stokes and pump beams at the radio frequencies f1 and f2 and a lock-in based demodulation of the CSRS signal. Taking advantage of CSRS' characteristic dependence on both excitation colors, the best fluoresence background suppression is obtained when demodulating the CSRS signal at f1-f2. Background-free LSM-CSRS imaging was demonstrated for samples of decreasing ratio of CSRS to fluoresence signal, namely: polymer beads, the epithelium and dermis of human skin, onion cells, avocado flesh and the wing disc of a Drosophila larva. Having removed the major obstacle for CSRS imaging, we introduced and quantified numerically the major interest of CSRS which is its unique backward radiation property in combination with high NA objective lenses. CSRS' backward radiation and its distinction from CARS is readily understood from the momentum conservation laws when considering all incident k-vectors forming the excitation focal spots. Using dynadic Green functions, we show numerically that the CSRS is predominantly forward directed for a homogeneous object, but the backward CSRS contribution rises to 1/4 for objects that are structured axially. Moreover, backward CSRS signal can even dominate forward CSRS (up to 100\%) if an annular Stokes illumination is applied. With an efficient Epi-CARS radiation at hand, various coherent Raman experiments become feasible which were impossible before. Just to name a few: Epi-detected confocal multi-focus CSRS; Epi-detected LSM-CSRS with a spectrometer at the descanned position; Epi-detected CSRS image scanning microscopy. Thus, we believe that this contribution is just the first milestone in CSRS microscopy with many others to follow.    

\section*{Methods: Experimental setup}
 A Yb-based fiber laser (APE Emerald engine, 80 MHz, 2–3 ps) is frequency doubled yielding 7~W of 515 nm output power. Parts of the emissions is used directly as Stokes beam to drive the CSRS process. The major part (4~W) of the 515~nm is employed to pump an optical parametric oscillator (OPO, APE Emerald). The OPO's signal beam is tunable to 660-950~nm and coupled into an external SHG unit. The latter generates up to 50~mW within the spectral range of 330-475~nm serving as pump beam for the CSRS four wave mixing. Thus, the 330-475~nm pump combined with the 515~nm Stokes beam allows addressing a Raman shift range from 1630-11000cm$^{-1}$. The pump and Stokes beams are superimposed in space and time via a dichroic beam splitter (Semrock, FF470-Di01-25x36) and a delay stage. Both beams are coupled into a home-built laser scanning microscope and focused by a 40x water objective lens (Nikon, Plan, NA = 1.15, immersion: water) into the sample. The excitation objective lens was replaced for a 60x objective (Nikon, Plan Apo TIRF, NA 1.45, immersion:oil) to generate the bead-oil interface image within Fig.~\ref{fig:experiment}. The CSRS radiation is collected by a condenser lens (Nikon, Achr-Apl, NA 1.4) in forward direction, spectrally separated from the broadband fluorescence background by means of 2 tilted bandpass filter (Semrock FF01-620/14-25 + FF01-605/15-25) and detected by a photo-electron multiplier (PMT, Thorlabs, PMT1001). For an enhanced suppression of the linear fluorescence background, 2 acousto-optic modulators (AOM, AA, MT200-A0.5-VIS) were applied to modulate the intensity of the Stokes and pump beams and at the frequencies f1~=~2.28~MHz and f2~=~3.75~MHz, respectively. The PMT output was demodulated simultaneously at the DC frequency, f1, f2 and at f1-f2~=~1.47~MHz using a lock-in amplifier (Zürich instruments, HF2LI). The lock-in time constant was set to 30~µs. All CSRS-images shown were recorded with a pixel dwell time of 40~µs.

\section*{Annex - numerical calculation}
\label{supinf}
\begin{wrapfigure}{r}{0.5\textwidth}
  \begin{center}    \fbox{\includegraphics[width=0.48\textwidth]{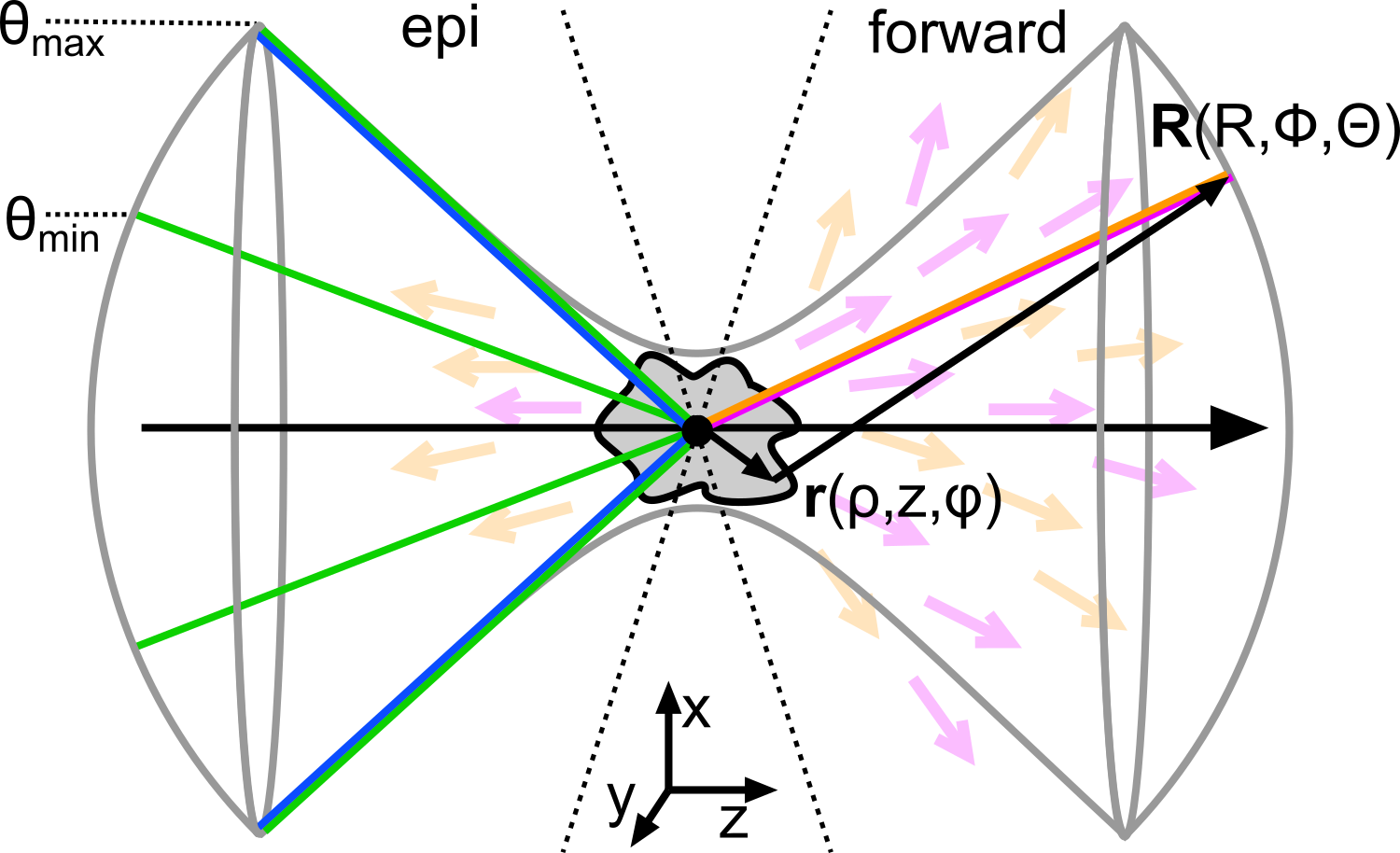}}
  \end{center}
  \caption{Declaration of variables}
    \label{Fig:coord}
\end{wrapfigure}

In the following, we shall summarize the equations used to generate Fig.~\ref{fig:object_support}b-e. The meaning of the variables is summarized in Fig.~\ref{Fig:coord}.\\
The focused field at the sample is given by the angular spectrum representation \cite{Cheng2002}:

\begin{equation}
\begin{tabular}[t]{rcl}
$\left[\begin{array}{r}
E_x(\rho,\phi,z) 	\\
E_y(\rho,\phi,z) 	\\
E_z(\rho,\phi,z) 	\\
\end{array}\right]$&
$ =\frac{ikf}{2}\exp(-ikf)$&
$\left[\begin{array}{c}
I_{00}+I_{02}\cos(2\phi)	\\
I_{02}\sin(2\phi)	\\
-i2I_{01}\cos(\phi) \\
\end{array}\right]$\\
\end{tabular}
\label{eq.:E_excitation_focus}
\end{equation}

Here $f$ denotes the focal length of the objective lens and the integrals $I_{0m}$ are provided by

\begin{equation}
I_{0m}=\int_{\theta_{min}}^{\theta_{max}}E_{inc}(\theta)\sin(\theta)[\cos(\theta)]^{1/2}g_m(\theta)\mathrm{J_m}[k\rho\sin(\theta)]\mathrm{d}\theta
\label{eq.:I_Integral}
\end{equation}

where $g_m$ equals $1+\cos(\theta)$, $\sin(\theta)$ and $1-\cos(\theta)$ for $m=0,1,2$, respectively. $J_m$ is the $m^{th}$ order Bessel function while $E_{inc}$ is the incoming electric field which we assumed to be x-polarized and constant within the (annular) aperture angles $\theta_{min}\leq \theta \leq\theta_{max}$.
The nonlinear polarization at anti-Stokes and coherent Stokes wavelength is given by: 

\begin{equation}
\begin{aligned}
P_{aS,a}^{(3)}(r)=3\chi_{abcd}^{(3)}(r)E_{p,b}E_{S,c}^*E_{p,d}\\
P_{cS,a}^{(3)}(r)=3\chi_{abcd}^{(3)}(r)E_{S,b}E_{p,c}^*E_{S,d}
\end{aligned}
\label{eq.:P_aS}
\end{equation}

Where a,b,c,d represent the polarization coordinates x, y or z. Using an x-polarized excitation, it was noticed that $\chi_{xxxx}^{(3)}$ dominates all other tensor components even under tight focusing conditions while filling the objective lens homogeneously \cite{Cheng2002}. Nevertheless, for the generation of Fig.~\ref{fig:object_support}c an annular mask with $\theta_{min}=56.5^\circ$ and $\theta_{max}=80^\circ$ was applied which does necessitate the inclusion of other tensor elements. 
For simplicity, we consider here only isotropic samples reducing the 81 susceptibility tensor elements to 21 which are nonzero \cite{Cheng2016}. Within isotropic media, these nonzero elements follow certain symmetry rules which are, $\chi_{1111}=\chi_{2222}=\chi_{3333}$, $\chi_{1122}=\chi_{1133}=\chi_{2211}=\chi_{2233}=\chi_{3311}=\chi_{3322}$, $\chi_{1212}=\chi_{1313}=\chi_{2323}=\chi_{2121}=\chi_{3131}=\chi_{3232}$, $\chi_{1221}=\chi_{1331}=\chi_{2112}=\chi_{2332}=\chi_{3113}=\chi_{3223}$. Further, it applies $\chi_{1111}=\chi_{1122}+\chi_{1212}+\chi_{1221}$ \cite{Cheng2016}. Within our simulations we were setting $\chi_{1122}=\chi_{1212}=\chi_{1221}=1$ and, hence, $\chi_{1111}=3$.
The nonlinear far-field radiation distributions is obtained using a dyadic Green function approach:

\begin{equation}
\begin{split}
&\left[\begin{array}{r}
E_{q,R}(R, \Theta, \Phi)\\
E_{q,\Theta}(R, \Theta, \Phi) 	\\
E_{q,\Phi}(R, \Theta, \Phi)	\\
\end{array}\right]
=-\frac{\omega_{q}^2}{c^2}\frac{\exp(ik_{q}\vert R\vert)}{\vert R\vert}
\iiint_{-\infty}^{\infty}\rho\mathrm{d}\rho\mathrm{d}\phi\mathrm{d}z\frac{\exp(ik_{q}\textbf{rR})}{\vert R\vert}\\ 
&\times\left[\begin{array}{r}
0\hspace{2.2 cm} 0 \hspace{1.8 cm} 0\hspace{0.5 cm}\\
\cos(\Theta) \cos(\Phi)\hspace{0.3 cm} \cos(\Theta)\sin(\Phi)\hspace{0.3 cm} -\sin(\Theta) 	\\
-\sin(\Phi) \hspace{1.25 cm}  \cos(\Phi) \hspace{1.3 cm}  0\hspace{0.5 cm} \\
\end{array}\right] 
\left[\begin{array}{r}
P_{q,x}^{(3)}(\textbf{r})\\
P_{q,y}^{(3)}(\textbf{r}) 	\\
P_{q,z}^{(3)}(\textbf{r})  \\
\end{array}\right]
\label{eq.:E_far_greens_function}
\end{split}
\end{equation}

where q is replaced by aS or cS to calculate either the anti-Stokes or coherent Stokes radiation. Within the simulations, we segmented the focal area into (121$\times$121$\times$121$\approx$) 1.77~mio elements of a width of 50~nm equally spaced into the x, y and z direction. The far-field radiation sphere was discretized into ($\Delta\Theta$=1$^{\circ}$, $\Delta\Phi$=2$^{\circ}$) 32400 elements. The coherent (anti-)Stokes radiation was qualified as either forward or backward directed if falling into the range $\Theta$.. 0-80$^{\circ}$ or $\Theta$.. 100-180$^{\circ}$, respectively.

\section*{Funding Information}

We acknowledge financial support from the Centre National de la Recherche Scientifique (CNRS), Aix-Marseille University (A-M-AAP-ID-17-13-170228-15.22-RIGNEAULT), A*Midex (ANR-11-IDEX-0001-02), Cancéropôle Provence-Alpes Côte d'Azur, French National Cancer institute (INCa), Région Sud, ANR grants (ANR-10-INSB-04-01, ANR-11-INSB-0006, ANR-16-CONV-0001), INSERM PC201508 and 18CP128-00.

\section*{Data availability}
The data that support the findings of this study are available from the corresponding author upon reasonable request.

\section*{Disclosure}
The authors declare no conflict of interest.

\bibliography{OSA-template}

\end{document}